# Fractal dimension and turbulence in Giant HII Regions

H. E. Caicedo-Ortiz[1], E. Santiago-Cortes[1,2], J. López-Bonilla[3], H.O. Castañeda[4]

[1] Maestría en Ciencias de la Complejidad, Universidad Autónoma de la Ciudad de México, San Lorenzo 290, esquina Roberto Gayol, Col. del Valle Sur, Del. Benito Juárez, C.P. 03100, México D.F.

[2] Grupo de Ingeniería y Tecnologías Cuánticas, Corporación Universitaria Autónoma del Cauca, Calle 5 No 3-85, Popayán, Colombia

[3] ESIME-Zacatenco, Instituto Politécnico Nacional, Edif. 5, CP 07738, México D.F.

[4] ESFM, Instituto Politécnico Nacional, Edif. 9, 1er piso, CP 07738, México D.F.

E-mail: hecaicedo@gmail.com

**Abstract.** We have measured the fractal dimensions of the Giant HII Regions Hubble X and Hubble V in NGC6822 using images obtained with the Hubble's Wide Field Planetary Camera 2 (WFPC2). These measures are associated with the turbulence observed in these regions, which is quantified through the velocity dispersion of emission lines in the visible. Our results suggest low turbulence behaviour.

## 1. Introduction

In the last years, the concept of fractals has been used intensively in the study of the Universe. The fractal dimension is employed as tool to study, for example, the solar flares [1], solar granulation [2], the morphology of craters on planets [3], galaxy catalogs [4], the study of the distribution of HII regions in galaxies [5] and the fractal dimension of the interstellar medium [6].

In this paper, we compute the fractal dimension of the Giant HII Regions of the Hubble V and Hubble X and we propose that the result can be interpreted as an indicator of the degree of hydrodynamic turbulence within the regions.

## 2. Observations

For the analysis of the fractal structure, we used images obtained by the Hubble Heritage Team. Hubble X observations were made on September the 7th, 1997 using F656N (H$\alpha$), F502N ([O III]) and F487N (H$\beta$ filters), with 2.6 hours of exposition time. On the other hand, Hubble V observations were performed on June the 18th in 1996 using F300W (U), F487N (H-$\beta$), F502N ([O III]), F547M (Strömgren y), F656N (H$\alpha$) and F658N ([N II]) filters and on August the 2nd in 1997 using the F555W (V) filter. The whole observations were processed using the WFPC tool of Hubble Special Telescope (HST). Figures 1 and 2 correspond to the Hubble X and Hubble V images used in this research.

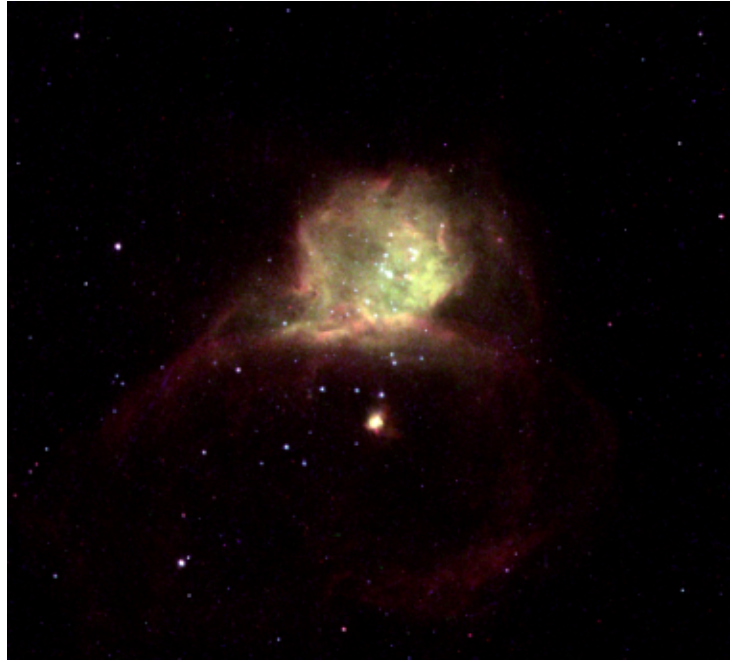

**Figure 1.** Hubble X. Hubble Heritage Team, NASA.

The spectra used in this study were obtained from data available at The Cambridge Astronomical Survey Unit (CASU) produced by the Isaac Newton Group of Telescopes located in the Canary Islands. The observations were carried out with the ISIS spectrograph of the William Herchel Observatory at the Roque de los Muchachos, using the technique of longslit spectroscopy, between August the $18^{th}$ and $19^{th}$ in 1992. The spectra for Hubble V (X) were obtained in 4 (8) different positions, all of them at position angle $90°$, with a slit width of 1" and a separation of 2" between the centers of two consecutive slit positions. Two spectra were taken simultaneously for each position, one in the range between 6390 Å and 6849 Å (red arm) and one between 4665 Å and 5065 Å (blue arm), both with a dispersion approximately 0.4 Å /pixel. The slit length was 200", with a spatial sampling along the slit of 0.34"/pixel in the red arm and 0.36"/pixel in the blue arm. Data reduction was performed using the IRAF data reduction software, following standard procedures.

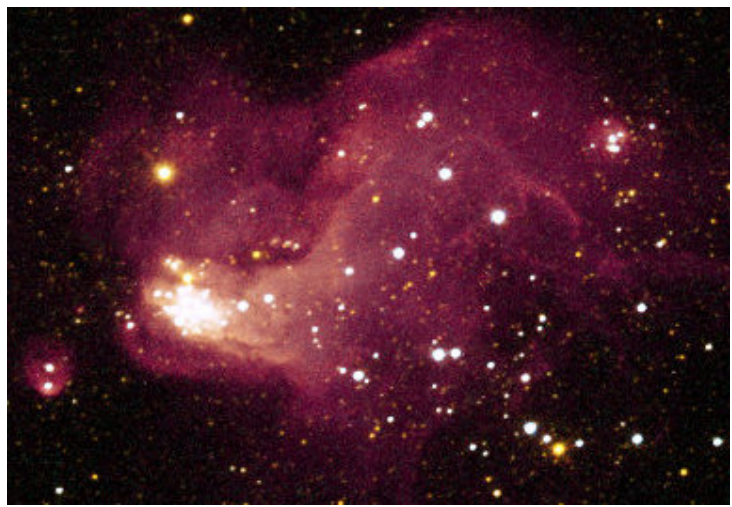

**Figure 2.** Hubble V. Hubble Heritage Team, NASA.

## 3. Fractal dimensión

The fractal structure is characterized by a parameter called fractal dimension, which is a generalization of the concept of Euclidean dimension. The more precise definition of the fractal dimension are in Hausdorff's work, become later known as Hausdorff dimensión [7]. This dimension is not practical in the sense that it is very difficult to compute even in elementary examples and nearly impossible to estimate in practical applications. There are several methods to define the fractal dimension [8].

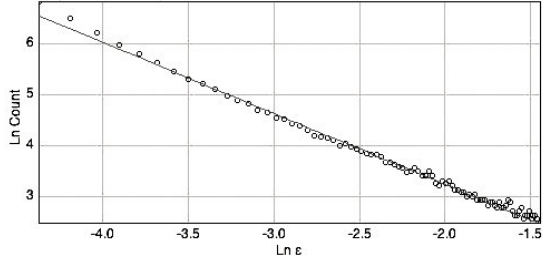
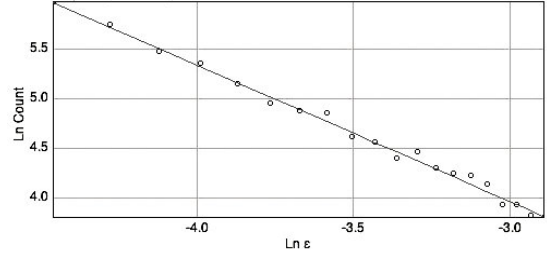

**Figure 3.** Fractal dimension of the perimeter of Hubble X.

**Figure 4.** Fractal dimension of the perimeter of Hubble V.

The most popular algorithm to compute the fractal dimension is the Box Counting ($D_B$) [8,9]. In this method, the object is covered with a set of boxes with side length ε. The number of $N(\varepsilon)$ boxes is proportional to scale ε in the form $N(\varepsilon) \propto 1/\varepsilon^{D_B}$, where the $D_B$ exponent is the object dimension. The dimension is defined by:

$$D_B = \lim_{\varepsilon \to 0} \frac{\log N(\varepsilon)}{\log(1/\varepsilon)}. \tag{1}$$

$N(\varepsilon)$ is the number of the boxes of side $\varepsilon$ necessary to cover the different points that have been registered in the physical space. As $\varepsilon \to 0$ then $N(\varepsilon)$ increases, $N$ meets the following relation:

$$N(\varepsilon) \cong k\varepsilon^{-D_B} \tag{2}$$

$$\log[N(\varepsilon)] \cong \log[k] - D_B \log[\varepsilon] \tag{3}$$

By means of least – square fitting of representation of $\log[N(\varepsilon)]$ versus $\log[\varepsilon]$, it has been obtained of the straight line regression given by equation (3), as is shown in Fig 3 and 4. The fractal dimension $D_B$ will come given by the slope of this equation. This methodology can be applied in systematic way because of the simplicity in the creation of the box covering and the computation is quickly performed, even for large images obtained with high resolution.

The fractal dimension of Hubble V and Hubble X was estimated using the automated dimension analysis plugin *Fraclac* of the images processing software *ImageJ*. A set of 12 measures was taken. Log-Log graphs were plotted of the reciprocal of the side length of the square against the number of out-lines containing squares (Fig 3 and 4). The fractal dimension for each HII region is the average of these 12 measures being 1.4081 for Hubble X and 1.3818 for Hubble V.

## 4. Velocity dispersion

The traditional method used to determine turbulence in a gaseous nebula is by velocity dispersion (σ) obtained from the width of the emission lines. The width of the emission lines measured in the observations presents excess after we take into considerations the contributions of the thermal movement of gas and the instrumental width. To estimate the instrumental width associated to every emission line, we used the emission lines observed in the calibration spectrum associated to each spectrograph arm (red and blue) and for each observed object (one for Hubble V, another for Hubble X, for arm). The instrumental width at half maximum (FWHM) associated to each emission line, is the width at the average medium height of the two emission lines on the calibration spectrum that narrow from upper to lower.

**Table 1.** Velocity dispersion for Hubble X and Hubble V with integrated spectral.

| Object | Hubble X | Hubble V | References |
|---|---|---|---|
| Hα | 10 ± 2 | 10 ± 2 | * |
| Hα | 10.5 ± 0.2 | 10.3 ± 0.3 | [10] |
| Hα | 10.4 ± 0.9 | 9.5 ± 0.8 | [11] |
| Hα | 9.8 | 9.5 | [12] |

* This research

We have that the component of instrumental width-associated velocity is expressed by $\sigma_{inst} = \dfrac{0.6006\, FWHM_{inst}}{1.414}$. $FWHM_{inst}$ is the instrumental width at half height expressed in velocity units. The velocity-associated to width observed in the emission line is expressed by $\sigma_{inst} = \dfrac{0.6006\, FWHM_{obs}}{1.414}$, with $FWHM_{obs}$ as the observed width (in velocity units). The thermal component is sensitive to the temperature assumed in the gas and to the electron-emitted mass in the emission lines. $\sigma_{inst} = \dfrac{166.3\, t}{1.414\, m}$, where $t = T_e \times 10^{-4}$, $T_e$ is the gas temperature, which in this case is 11500 K [12] and $m$ is the atomic mass of the ion related to the emission line. The velocity dispersion is:

$$\sigma_{disp} = \left(\sigma_{obs}^2 - \sigma_{inst}^2 - \sigma_{term}^2\right)^{1/2}. \tag{4}$$

Table 1 contains the results of the velocity dispersion in Hα for Hubble X and Hubble V for the integrated spectra for this research and by other previous research.

## 5. Discussion

The origin of turbulence cannot be easily determined, but it is know that both the dynamic and the thermal effects contribute strongly to turbulence. In the last years, the fractal theory has been applied at the study of fluid dynamics. These properties are present in fluids that show the transition from a laminar structure to a turbulent regime [14]. Sreenivasan and Meneveau [15] introduced the fractal theory into experimental fluid dynamics, and analyzed the fractal dimension of a low speed shear layer, jet, wake and boundary layer. Sreenivasan discussed the fractal and multifractal description and measurement of turbulent flow [16].

The value of the fractal dimension of 1.4081 for Hubble X and 1.3818 for Hubble V suggest a fractal behavior of these giant HIII regions. From the geometric point of view, the fractal dimension is a measure of the system irregularity. As the fractal dimension increases, the degree of irregularity of the curve is higher. Zhao YuXin et al [14], suggests that the transition from a laminar to a turbulent flow can be seeing as a process of breaking up of the streamline and contour. The streamline and contour of laminar flows are smooth curves. When the flow tends to be instable, they will curl and twist. Vortex of different scales will interact and transfer energy, and while the flow make the transition to turbulence, the streamline and contour will break up completely and it will show irregularity of the transitional mixing layer, which can be evaluated by the fractal dimension [14]. Table 2 presents the fractal dimension of different fractal objects, of Hubble X, Hubble V and another HIII regions for whose this property has been calculated. The few existent studies about fractal dimension made to HIII regions [17-20] only calculate it and suggest the possibility of associate the behavior of this HIII regions with another fractal structures.

The Hubble X and Hubble V dimension fractal values suggest a behavior slightly irregular that can be interpreted as a low turbulence system. To check this hypothesis is necessary to determine the turbulence in Hubble X and Hubble V.

**Table 2.** Fractal dimension of different fractal objects and HII regions.

| Object | $D_B$ | Reference | Object | $D_B$ | Reference |
|---|---|---|---|---|---|
| Vela nebula | $1.2 \pm 0.05$ | [16] | Paley I | 1.677181 | [18] |
| Eta Carinae nebula | $1.37 \pm 0.02$ | [16] | Rosette Nebulae (close-up) | 1.78261 | [18] |
| Hubble V | 1.3818 | * | Eagle Nebulae (close-up) | 1.669622 | [18] |
| Hubble X | 1.4081 | * | IC434 (col I) | 1.811718 | [19] |
| Julia set | 1.679594 | [17] | Helix Nebula (NGC 7293) | 1.841059 | [20] |
| Horsehead nebula | 1.6965725 | [17] | | | |

* This research

As stated previously, the velocity dispersion is one of the methods used to measure turbulence in HII regions. The natural width is a diagnosis method of a turbulence field of velocity.

In order for HII regions to have a supersonic turbulence movement is necessary the continuous input of energy to replace the dissipated energy due to the viscosity of the gas and shocks. The first source of energy is radiation from ionizing stars. Another source that can produce supersonic movement of gases in HIII regions are the stellar winds. O and B stars inject about $10^{-(5-6)}$ $M_\odot$ per year with a velocity that exceeds the 1000 km/s. Considering that HIII regions are ionized by a large number of high mass stars ($10^2$ - $10^3$), their stellar winds affect the nebular gas cinematic and the kinetic energy associated to the winds could produce a considerable turbulence. A third mechanism associated to this kind of turbulent movements is gravity.

Giant HIII regions are ionized by a large number of massive stars. It is possible estimate to the total mass of the regions if we use a typical mass function for the ionizing clusters is expected and the neutral and ionized gas and ionized gas is included, allowing to estimate masses of HIII regions higher than $10^6$ $M_\odot$ [21,22].

Our results for Hubble X and Hubble V show that the ionized hydrogen is described by subsonic behavior, with velocities lower than sound (of the order of 10 km/s). The dispersion velocity values are in the limit of supersonic-subsonic. Both HII regions exhibit a turbulent behavior that is located at the transition subsonic-supersonic, which is a very peculiar case for typical giant HII regions, because of for these types of astronomical objects; the nebular gas presents dispersion velocities on the average of 25 km/s.

The results obtained through the computation of the fractal dimension of Hubble X and Hubble V are compatible with the turbulent behavior of the ionized gas of these HII regions. The principal result of this research is to focus on the meaning of the fractal dimension in these Giant HII regions as an element that allows characterizing their turbulent properties. This interpretation gives a new tool and methodology for the future study of turbulence in other Giant HII Regions.

**References**
[1] Aschwanden M J and Aschwanden P D 2008 *AJ* **674** 530, Aschwanden M J and Stern R A and Güdel M 2008 *AJ* **672** 659
[2] Greimel R, Brandt P N, Guenther N and Mattig W 1990 *Vistas in Astronomy* **3** 413
[3] Ching D, Taylor G J, Mouginis-Mark P and Bruno B C 1993 *Lunar Planet Sci* **XXIV** 283
[4] Thanki S, Rhee G and Lepp S 2009 *AJ* **138** 941
[5] Sanchez N and Alfaro E J 2008 *ApJS* **178** 1
[6] Sanchez N, Alfaro E J and Pérez E 2009 *RevMexAA Conf. Ser.* **35** 76
[7] Hausdorff F 1918 *Mass. Mathematische Ann.* **79** 157-179
[8] Peitgen H O, Jurgens H, and Saupe D 2004 *Chaos and Fractals - New Frontiers of Science* (New York: Springer-Verlag)
[9] Theiler J 1990 *J. Opt. Soc. Am. A* **7**(6) 1055
[10] Hippelein H H 1986 *A&A* **160** 374
[11] Melnick J, Moles M, Terlevich R and Garcia-Pelayo J M 1987 *MNRAS* **226** 849
[12] Roy J R, Arsenault R and Joncas G 1986 *ApJ* **300** 624
[13] Peimbert A, Peimbert M and Ruiz M T 2005 *ApJ* **634** 1056
[14] Zhao Y X, Yi S, Tian L F, He L and Cheng Z Y 2008 *Sci China Ser G-Phys Mech Astron* **51** 8 1134
[15] Sreenivasan K R and Meneveau C 1986 *J Fluid Mech* **173** 357
[16] Sreenivasan K R 1991 *Ann Rev Fluid Mech* **23** 539
[17] Blancher S and Perdang J 1990 *Vistas in Astronomy* **33** 393
[18] Datta S 2001 Fractal structure of molecular clouds Preprint astro-ph/0105036
[19] Datta S 2003 *A&A* **401** 193
[20] Datta S 2003 *Planetary Nebulae: Their Evolution and Role in the Universe* vol 209 (France: Astronomical Society of the Pacific) p 523
[21] Terlevich R and Melnick J 1981 *MNRAS* **195** 839
[22] Kennicutt R C Jr 1984 *ApJ* **287** 116